\documentclass[twocolumn]{aastex63}

\received{October 20, 2022}
\revised{November 28, 2022}
\accepted{\today}
\submitjournal{ApJ}

\pdfoutput=1 
\usepackage{amsmath,amstext}
\usepackage[T1]{fontenc}
\usepackage[figure,figure*]{hypcap}

\usepackage{graphicx}
\usepackage{graphics}
\usepackage{color}
\usepackage[caption=false]{subfig}
\definecolor{dartmouthgreen}{rgb}{0.05, 0.5, 0.06}
\usepackage{afterpage}
\usepackage{natbib}
\usepackage[percent]{overpic}

\def\arcsec{\hbox{$^{\prime\prime}$}}
\def\deg{\hbox{$^\circ$}}
\def\hr{\textsuperscript{h}}
\def\min{\textsuperscript{m}}
\def\sec{\textsuperscript{s}\hspace{-0.7mm}}

\def\nh{$N_{\rm H}$}
\def\nhm{\textit{N}_{\rm{H}}/\rm{cm}^{-2}}
\def\lx{$L_{2-10\ \rm{keV}}$}

\def\lfour{$L_{22\mu \rm{m}}$}
\def\ltwo{$L_{4.6\mu \rm{m}}$}
\def\ltwel{$L_{12\mu \rm{m}}$}

\def\nh{$N_\mathrm{H}$}

\def\chandra{\textit{Chandra}}
\def\xmm{\textit{XMM-Newton}}
\def\wise{\textit{WISE}}
\def\nustar{\textit{NuSTAR}}
\def\swift{\textit{Swift}/BAT}

\def\ngc4178{NGC\,4178}

\graphicspath{ {Images/} }

\shorttitle{\nustar{} Observations of \ngc4178{} and J0851+3926}
\shortauthors{Pfeifle et al.}
\begin{document}
\title{\nustar{} Observes Two Bulgeless Galaxies: No Hard X-ray AGN Detected in \ngc4178{} or J0851+3926}

\correspondingauthor{Ryan W. Pfeifle}
\email{ryan.w.pfeifle@nasa.gov}

\author[0000-0001-8640-8522]{Ryan W. Pfeifle}
\altaffiliation{NASA Postdoctoral Program Fellow}
\affiliation{X-ray Astrophysics Laboratory, NASA Goddard Space Flight Center, Code 662, Greenbelt, MD 20771, USA}
\affiliation{Oak Ridge Associated Universities, NASA NPP Program, Oak Ridge, TN 37831, USA}

\author[0000-0003-2277-2354]{Shobita Satyapal}
\affiliation{Department of Physics and Astronomy, George Mason University, 4400 University Drive, MSN 3F3, Fairfax, VA 22030, USA}

\author[0000-0001-5231-2645]{Claudio Ricci}
\affiliation{N\'ucleo de Astronom\'ia de la Facultad de Ingenier\'ia, Universidad Diego Portales, Av. Ej\'ercito Libertador 441, Santiago, Chile}
\affiliation{Kavli Institute for Astronomy and Astrophysics, Peking University, Beijing 100871, China}
\affiliation{Department of Physics and Astronomy, George Mason University, 4400 University Drive, MSN 3F3, Fairfax, VA 22030, USA}

\author[0000-0002-4902-8077]{Nathan J. Secrest}
\affiliation{U.S. Naval Observatory, 3450 Massachusetts Avenue NW, Washington, DC 20392, USA}

\author[0000-0002-8818-9009]{Mario Gliozzi}
\affiliation{Department of Physics and Astronomy, George Mason University, 4400 University Drive, MSN 3F3, Fairfax, VA 22030, USA}

\author[0000-0002-4375-254X]{Thomas Bohn}
\affiliation{Hiroshima Astrophysical Science Center, Hiroshima University, 1-3-1 Kagamiyama, Higashi-Hiroshima, Hiroshima 739-8526, Japan}

\author[0000-0003-4693-6157]{Gabriela Canalizo}
\affiliation{University of California, Riverside, Department of Physics \& Astronomy, 900 University Ave., Riverside, CA 92521}

\author[0000-0003-4701-8497]{Michael A. Reefe}
\altaffiliation{National Science Foundation, Graduate Research Fellow}
\affiliation{Department of Physics and Astronomy, George Mason University, 4400 University Drive, MSN 3F3, Fairfax, VA 22030, USA}
\affiliation{Department of Physics, Massachusetts Institute of Technology, Cambridge, MA 02139}
\affiliation{Kavli Institute for Astrophysics and Space Research, Massachusetts Institute of Technology, Cambridge, MA 02139}

\begin{abstract}
The discovery over the last several decades of moderate luminosity AGNs in disk-dominated galaxies -- which show no ``classical'' bulges -- suggests that secular mechanisms represent an important growth pathway for supermassive black holes in these systems. We  present new follow-up \nustar{} observations of the optically-elusive AGNs in two bulgeless galaxies, \ngc4178{} and J0851+3926. \ngc4178{} was originally reported as hosting an AGN based on the detection of [Ne V] mid-infrared emission detected by \textit{Spitzer}, and based on \chandra{} X-ray imaging it has since been argued to host either a heavily obscured AGN or a supernova remnant. J0851+3926 was originally identified as an AGN based on its \wise{} mid-IR colors, and follow-up near-infrared spectroscopy previously revealed a hidden broad line region, offering compelling evidence for an optically-elusive AGN. Neither AGN is detected within the new \nustar{} imaging, and we derive upper limits on the hard X-ray 10-24 keV fluxes of $<7.41\times10^{-14}$\,erg\,cm$^{-2}$\,s$^{-1}$ and $<9.40\times10^{-14}$\,erg\,cm$^{-2}$\,s$^{-1}$ for the AGNs in \ngc4178{} and J0851+3926, respectively. If these non-detections are due to large absorbing columns along the line of sight, the non-detections in \ngc4178{} and J0851+3926 could be explained with column densities of log($N_{\rm{H}}/\rm{cm}^2)>24.2$ and log($N_{\rm{H}}/\rm{cm}^2)>24.1$, respectively. The nature of the nuclear activity in \ngc4178{} remains inconclusive; it is plausible that the [Ne V] traces a period of higher activity in the past, but that the AGN is relatively quiescent now. The non-detection in J0851+3926 and multiwavelength properties are consistent with the AGN being heavily obscured.
\end{abstract}

\keywords{Active galactic nuclei --- 
AGN host galaxies  --- Low-luminosity active galactic nuclei}

\section{Introduction} 
\label{sec:intro}
The well-known correlation between the black hole mass and the host galaxy's stellar velocity dispersion \citep[e.g.,][]{magorrian1998, gebhardt2000, gultekin2009, mcconnell2013} launched a long-standing view that black hole growth and the build-up of galaxy bulges go hand-in-hand, perhaps as interactions fuel the central SMBH and grow the bulge, and feedback from the active galactic nucleus (AGN) regulates the surrounding star formation in the host galaxy \citep[e.g.,][]{toomre1977,barnes1992,silk1998,kauffmann2000,hopkins2012}.  The connection between black hole growth and galaxy bulges was further intimated by the fact that until recently, virtually all currently known actively accreting black holes - i.e. AGNs - in the local Universe were found in galaxies with prominent bulges \citep[e.g.,][]{ho1997,kauffmann2003}. In recent years, however, it has now become clear that many moderate luminosity ($L_{0.5-8\,\rm{keV}}=10^{42}$-$10^{44}$\,erg\,s$^{-1}$) AGNs are found in disk-dominated galaxies with no signs of interactions \citep[e.g.,][]{simmons2012, kocevski2012}. Although some mergers can regrow or leave large scale disks \citep{springel2005,hopkins2009}, and it may be possible to build a ``classical'' bulge without mergers \citep{bell2017}, it is extremely likely that the vast majority of disk-dominated galaxies in the local Universe have evolved largely independently from mergers since z $\approx$ 2 \citep{martig2012}. The discovery of significant AGN activity in disk galaxies therefore stresses the importance of secular pathways in black hole growth. Surprisingly, in the few disk-dominated galaxies with optically identified AGNs, the black hole mass is found to correlate with the total stellar mass of the disk \citep[e.g.,][]{cisternas2011,simmons2017}. Since the bulge carries the imprint of the merger history of a galaxy, this suggests that SMBH and galaxy co-evolution takes place largely independently from mergers. However, most past studies are based on optical spectroscopic observations, which can be severely limited in the study of bulgeless galaxies.  It is well known that the infrared-to-blue luminosity ratio observed in galaxies increases along the Hubble Sequence, implying that late-type galaxies are extremely dusty \citep[e.g.,][]{deJong1984}. In addition, the optical emission lines used to identify AGNs get increasingly diluted by star formation in the host as the bulge to disk ratio decreases \citep[e.g.,][]{trump2015}. In an effort to study \textit{obscured} AGN growth in bulgeless galaxies, we focus in this work on \ngc4178{} and SDSS\,J085153.64+392611.76 (hereafter J0851+3926), both of which have been reported to host optically elusive, heavily obscured AGNs.

\ngc4178{} is a bulgeless spiral galaxy residing at a distance of 16.2 Mpc away \citep{kent2008} in the Virgo Cluster. \citet{satyapal2009} first reported on the presence of an AGN in \ngc4178{} based on the detection of a high-ionization [Ne V]$\lambda 14.3\,\rm{\mu m}$ emission line -- a reliable tracer of AGN activity \citep{abel2008} in the mid-infrared (mid-IR) -- using the \textit{Spitzer Space Telescope}. Given that the optical emission shows no sign of an AGN and instead is consistent with an HII star forming region \citep[e.g.,][]{secrest2012}, the detection of the [Ne V] emission line suggested that \ngc4178{} in fact hosted an optically elusive AGN. Follow-up \chandra{} observations presented in \citet{secrest2012} revealed a weak ($5.3\sigma$) and predominantly soft X-ray point source coincident with the nucleus of the galaxy. \citet{secrest2012} concluded that the X-ray source properties were consistent with a heavily obscured AGN with an absorbing column density of $N_{\rm{H}}=5\times10^{24}$\,cm$^{-2}$, covering factor of C=0.99, and a photon index of $\Gamma=2.3^{+0.6}_{-0.5}$. Using (1) the [Ne V] emission line flux to estimate the bolometric luminosity, as well as (2) an archival VLA image to utilize the fundamental plane of black hole accretion \citep[e.g.,][]{gultekin2009b}, \citet{secrest2012} estimated that the AGN is powered by a $\sim7\times10^4-2\times10^5\,\rm{M}_{\odot}$ intermediate mass black hole. However, the AGN interpretation has recently been called into question by \citet{hebbar2019}, who instead claimed that the X-ray emission is better fit by a hot plasma model and that the X-ray emission is likely due to a supernova remnant. \citet{graham2019}, too, recently questioned the AGN interpretation, suggesting that perhaps the [Ne V] emission traces a period of previous activity but that the AGN is relatively quiescent now.

J0851+3926 is a bulgeless spiral galaxy at $z=0.1296$ originally selected by \citet{satyapal2014} based on its Wide-Field Infrared Survey Explorer (\textit{WISE}) mid-IR colors; J0851+3926 satisfied the stringent 3-band mid-IR AGN color cut defined by \citet{jarrett2011}, suggestive of a powerful, dust obscured AGN. J0851+3926 was selected again in \citet{bohn2020} in an effort to measure black hole masses in optically elusive AGNs in bulgeless galaxies. While the Baldwin-Phillips-Terlevich \citep[BPT,][]{baldwin1981} optical spectroscopic line ratios are consistent with a composite galaxy \citep{bohn2020} -- which are commonly assumed to arise from a mixture of AGN and star formation driven emission, based on the criteria laid out in \citet{kewley2001,kewley2006,kauffmann2003} -- shocks can also give rise to similar emission line ratios, leading to an ambiguous optical classification \cite[e.g.,][]{allen2008,rich2011,rich2014}. Coupled with the lack of Balmer emission lines in the optical band, there is no definitive evidence in the optical for an AGN \citep{satyapal2014,bohn2020}. As a part of their elusive AGN campaign, \citet{bohn2020} published near-infrared spectra obtained from the Near-Infrared Echellette Spectrometer (NIRES) and Near-Infrared Spectrometer (NIRSPEC) on Keck for J0851+3926 and reported the detection of a broad ($1489\pm184$\,km\,s$^{-1}$ in NIRSPEC, $1363\pm31$\,km\,s$^{-1}$ in NIRES) Pa$\alpha$ emission line in both observations. After ruling out the possibility that the broad emission line is due to outflows or a supernova, the authors attribute the broad Pa$\alpha$ emission to an optically elusive AGN. An X-ray AGN was not detected in the \chandra{} imaging, but the $3\sigma$ upper limit on the 2-10\,keV flux implied a column density of log($N_{\rm{H}}/\rm{cm}^{2})\geq24.43$ \citep{bohn2020} based on the relationship between the observed 2-10\,keV and 12\,$\rm{\mu m}$ emission derived in \citet{pfeifle2022}. This column density is consistent with the lack of broad optical emission lines and the derived level of extinction \citep[$E_{\rm{H\alpha}}$($B-V \geq1.40$)][]{bohn2020}. Virial mass measurements using the broad Pa$\alpha$ emission yielded an (extinction corrected) mass of log(M/$\rm{M}_{\odot})=6.78\pm0.50$ for the optically elusive, heavily obscured AGN \citep{bohn2020}. 

Here we report on new \nustar{} observations of \ngc4178{} and J0851+3926 which were carried out with the goal of understanding whether AGNs are present in these galaxies and to provide constraints on the obscuring columns. We organize this work as follows: in  Section~\ref{sec:reduction} we describe the observations and processing of the \nustar{} and \chandra{} observations, and in Section~\ref{sec:analysis} we describe our data analysis. In Section~\ref{sec:results} we describe our results. In Section~\ref{sec:discussion} we discuss our work in the context of our previous work and the literature, and in Section~\ref{sec:conclusion} we present our conclusions. Throughout this manuscript we adopt the following cosmology: $\textrm{H}_0 = 70$\,km\,s$^{-1}$\,Mpc$^{-1}$, $\Omega_M=0.3$, and $\Omega_\Lambda=0.7$.

\section{Observations and Data Processing}
\label{sec:reduction}

\subsection{\nustar{} Observations}
\nustar{} observations for \ngc4178{} (58.3 ks Obs. ID 60601013002, PI Satyapal) and J0851+3926 (61.4 ks, Obs. ID 60501025002, PI Satyapal) were conducted on 25 June 2020 and 08 May 2020, respectively (see Table~\ref{table:nustarobs}). Reprocessing of the data was performed using the \nustar{} Data Analysis Software \citep[\textsc{nustardas},][]{nustardas}\footnote{https://heasarc.gsfc.nasa.gov/docs/nustar/analysis/} v0.4.7 package available in \textsc{heasoft} v6.27 \citep{heasoft} along with the latest \textsc{caldb} and clock correction file at that time. The \textsc{nupipeline} script was used to conduct Stage 1 and 2 reprocessing, and we used the background light curves for each observation to inform our choices for the South Atlantic Anomaly (SAA) calculation. For both observations we used SAA algorithm 1, we chose the `Optimized' and `Strict' calculation modes for \ngc4178{} and J0851+3926, respectively, and we did not employ the `tentacle' option\footnote{See page 35 of the \textsc{nustardas} data analysis software guide for a description of the `tentacle' option.} in either case since a stable background light curve was achieved using the aforementioned algorithm settings. Energy filtered event files were created using \textsc{dmcopy} in the Chandra Interactive Analysis of Observations (\textsc{ciao}) software package \citep{fruscione2006}\footnote{A note of caution: For the filtered exposure time, \textsc{dmextract} searches the header for the keyword `DTCOR', but \nustar{} observations use the keyword `DTIME'. To retrieve the correct exposure time alongside the source counts, one needs to modify the event file header to include `DTCOR'. Otherwise, \textsc{dmextract} assumes that the exposure time is equal to the live time.}. The \textsc{nuproducts} script was used to extract the Stage 3 spectroscopic data products using the source and background regions described in Section~\ref{sec:analysis}. Due to the low number of counts, extracted spectra were grouped by 1\,count per bin using the \textsc{heasoft} \textsc{grppha} command in order to use Cash statistics \citep{cash1979} during fitting. 

\subsection{\chandra{} Observation}
In addition to the \nustar{} observations of \ngc4178{} and J0851+3926, we also retrieved their archival \chandra{} observations. \ngc4178{} was observed for 40\,ks (Observation ID\,12748) on 19 February 2011 and examined in both \citet{secrest2012}, \citet{hebbar2019}, and mostly recently in \citet{graham2019}. J0851+3926 was observed for 20\,ks (Observation ID 22584) and was examined in \citet{bohn2020}. These observations were reprocessed using \textsc{ciao} v.4.12 and \textsc{caldb} v4.9.1 using the \textsc{chandra\_repro} script. Energy filtered images were again created using \textsc{dmcopy} within \textsc{ciao}. Spectra were extracted using the \textsc{specextract} script and, as in the case of the \nustar{} data, were grouped by 1 count per bin.

\begin{table}[t]
\begin{center}
\caption{NuSTAR Observations}
\label{table:nustarobs}
\begin{tabular}{ccccc} 
\hline
\hline
\noalign{\smallskip}
\noalign{\smallskip}
Name & ObsID & FPM & Exp. & Net Exp. \\
 &  &  & (ks) & (ks) \\
 (1) & (2) & (3) & (4) & (5) \\
\noalign{\smallskip}
\noalign{\smallskip}
\hline
\noalign{\smallskip}
  \ngc4178{} & 60601013002 & A & 58.3 & 54.0 \\ 
  & 60601013002 & B & 58.3 & 53.4 \\ 
\noalign{\smallskip}
J0851+3926 & 60501025002 & A & 61.4 & 52.9 \\ 
 & 60501025002 & B & 61.4 & 52.5 \\ 
\noalign{\smallskip}
\hline
\end{tabular}
\end{center}
\tablecomments{Relevant information for \nustar{} data observations. Columns 1-3: target name, observation ID, and camera. Columns 4-5: total, raw exposure time and net exposure time after processing the data.}
\end{table}

\section{Analysis}
\label{sec:analysis}

\subsection{\nustar{} Source Detection and Photometry}
Neither \ngc4178{} nor J0851+3926 host an obvious hard X-ray point source upon visual inspection. To search the \nustar{} FPMA and FPMB images for potential sources, we employed \textsc{wavdetect} from \textsc{ciao} on the observed frame 3$-$8\,keV band, as well as the more traditional 3$-$10\,keV, 10$-$24\,keV, and 3$-$24\,keV \nustar{} bands for both galaxies. \textsc{wavdetect} found no sources in either FPMA or FPMB at the location of J0851+3926. One source is detected by \textsc{wavdetect} in the 3$-$8\,keV, 3$-$10\,keV, and 3$-$24\,keV FPMA images of NGC\,4178, but the source is not found in the FPMA 10$-$24\,keV band nor is it detected in FPMB in any band. Because the detections in the 3$-$10\,keV and 3$-$24\,keV bands are due largely to the detection of a source in the 3$-$8\,keV band, and because the \textsc{wavdetect} 3-10 keV region was not well centered on the source centroid, we used the 3$-$8 keV band for the source aperture placement for \ngc4178{}.

Since \textsc{wavdetect} did not find any source at the location of J0851+3926, we instead used an aperture of 45\arcsec{} in radius centered on the SDSS optical position of the galaxy to extract counts from both FPMA and FPMB. An annular region with an inner and outer radius of 90\arcsec{} and 150\arcsec{}, respectively, was also centered on the galaxy's position and used to sample the background for both cameras. 

The placement of apertures for source and background extraction for \ngc4178{} required more care. Although the average astrometry accuracy of \nustar{}'s pointing is often quoted as $\sim8$\arcsec{} for bright sources \citep{harrison2013}, the average positional accuracy of \nustar{} can range from $\sim14$\arcsec{} for the brightest sources to $\sim22$\arcsec{} for the weakest sources \citep[see, e.g., Figure~4 and Section~3.1 in ][]{lansbury2017}. The source identified by \textsc{wavdetect} in the FPMA 3-8 keV band located at $\alpha$ = 12\hr{}12\min{}44.\sec{}86, $\delta$ = +10\deg{}51\min{}20.\sec{}71 (in FPMA) lies $\sim40.1$\arcsec{} from the putative AGN's position \citep[$\alpha$ = 12\hr{}12\min{}46.\sec{}32, $\delta$ = 10\hr{}51\min{}54.\sec{}61, derived from \chandra{} X-ray imaging,][]{secrest2012}. Given the average range of offsets observed by \citet{lansbury2017}, if this source is indeed the AGN, its positional offset from the reported \chandra{} position would be exceptionally large. On the other hand, the source is offset by only $\sim8.8$\arcsec{} from the ultraluminous X-ray source (ULX), located at $\alpha$ = 12\hr{}12\min{}44.\sec{}51,
$\delta$ = 10\deg{}51\min{}13.\sec{}64 within \ngc4178{} \citep{secrest2012}, suggesting that the source detected by \nustar{} is in fact the ULX. If we assume that the positional offset between the \nustar{} and \chandra{} positions is due strictly to the positional accuracy of \nustar{}, the source's positional offset from the AGN is $\sim49$\arcsec{}, exactly that found for the ULX \citep{secrest2012}. We therefore conclude that the source detected by \nustar{} is more likely the ULX. 

\begin{figure}
    \centering
    \includegraphics[width=0.99\linewidth]{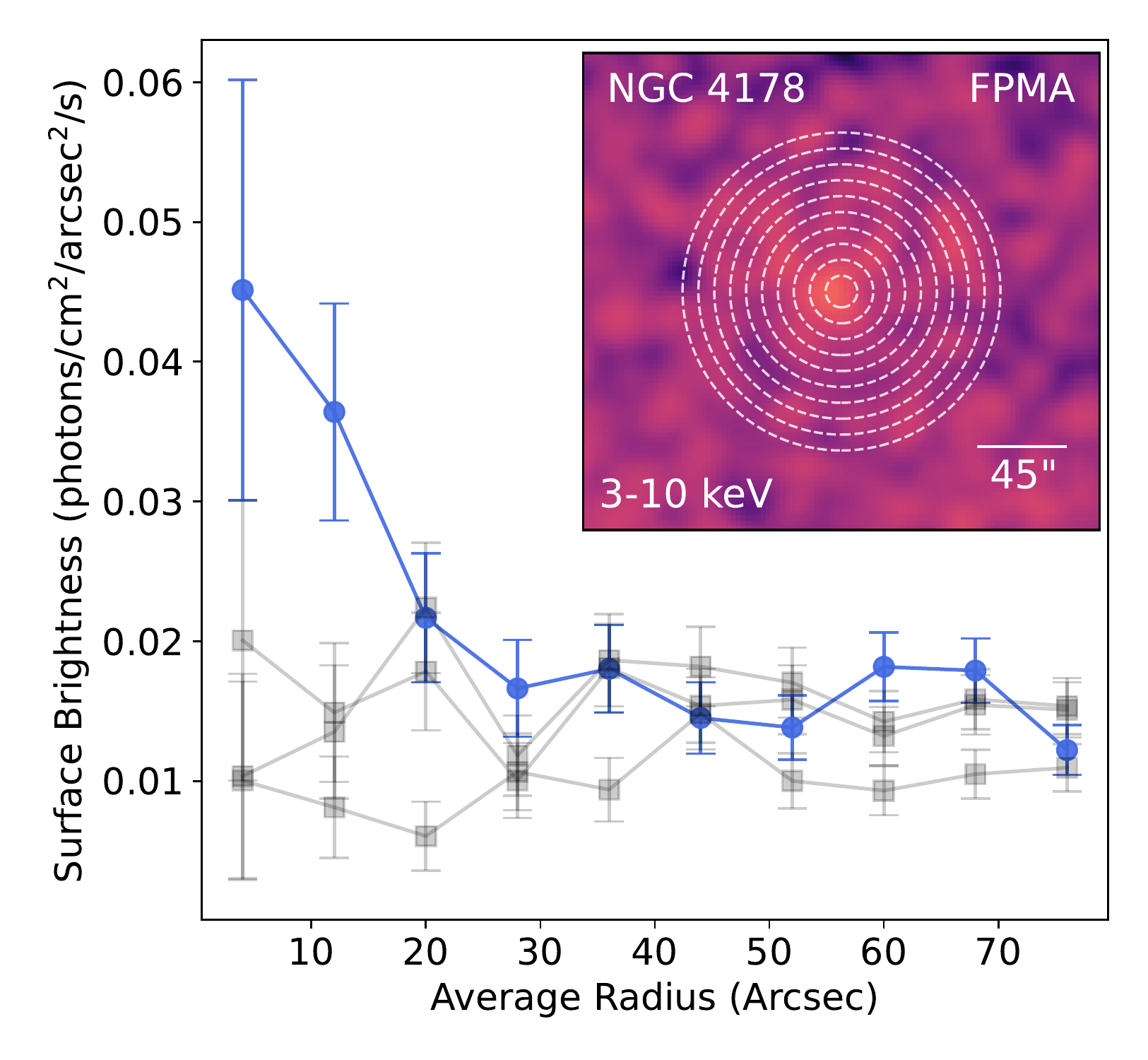}
    \caption{Surface brightness as a function of radial distance from the ULX in NGC\,4178. The source in the \nustar{} FPMA 3$-$10~keV band is shown in the top right corner, where the data are smoothed using a 3-pixel Guassian kernel and displayed with a perceptually uniform color map. Concentric, white, dashed circles indicate the annuli from which counts were extracted to construct the radial profile of the ULX. The radial profile is shown using the blue points connected with a line; error bars represent the standard error. The surface brightness of the source decreases from the source center out to $\sim28$\arcsec{}, after which the surface brightness varies little with radial distance. Radial surface brightness profiles for three nearby, background-dominated areas are shown in gray for comparison.  
    }
    \label{fig:surfbrght}
\end{figure}

\begin{figure*}[ht!]
    \centering
    \subfloat{\includegraphics[width=0.99\textwidth]{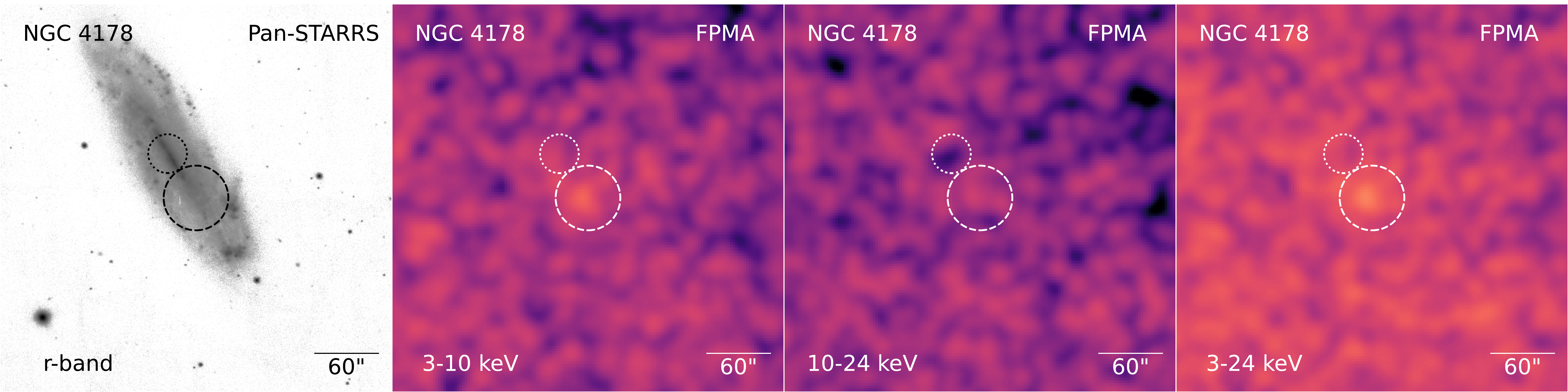}}\\
    \vspace{-4mm}
    \subfloat{\includegraphics[width=0.99\textwidth]{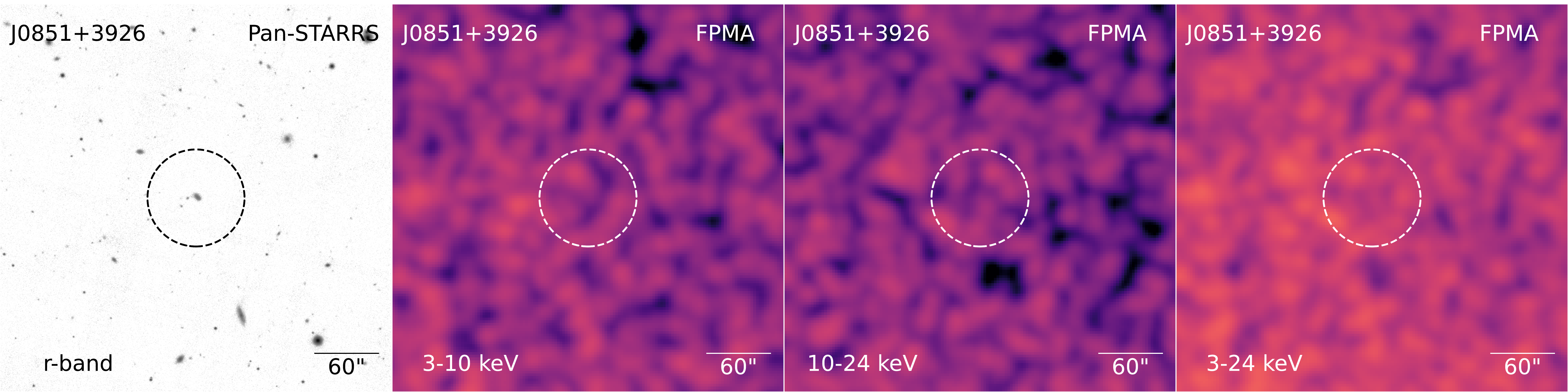}}\\
    \caption{Pan-STARRS r-band and \nustar{} FPMA imaging for \ngc4178{} (top) and J0851+3926 (bottom). Left to right: Pan-STARRS r-band, \nustar{} FPMA 3-10 keV, 10-24 keV, and 3-24 keV bands. X-ray images are smoothed using a three-pixel Gaussian kernel and displayed with the perceptually uniform sequential color map `magma' in \textsc{matplotlib}. Top: dashed $30$\arcsec{} radius circles represent the extraction region for the ULX, while dotted $18$\arcsec{} radius circles represent the AGN extraction region; these circles are offset by $\sim49$\arcsec{} from one another. Bottom: dashed $45$\arcsec{} radius circles represent the AGN extraction region. 
    }
    \label{fig:imaging}
\end{figure*}

Due to the angular resolution of \nustar{} \citep[$\sim$18\arcsec{} FWHM,][]{harrison2013} and the angular separation between the ULX and putative AGN ($\sim$49\arcsec{}), the source apertures had to be chosen carefully so as to avoid (as much as possible) contamination at the position of the putative AGN by the nearby, brighter ULX. This is particularly important in the calculation of upper limits in the event of a non-detection for the AGN. To better understand the radial extent of the ULX, we placed a series of equally-spaced concentric annuli with radii extending from 0-80\arcsec{} and measured the number of counts within each annulus. We show the radial profile for the ULX along with the \nustar{} 3-10 keV image with the annuli overlaid in Figure~\ref{fig:surfbrght}. The surface brightness decreases steadily from 0\arcsec{} out to $\sim$28\arcsec{} from the ULX, beyond which the surface brightness becomes fairly constant. We can therefore justify the choice of a 30\arcsec{} radius aperture centered on the position of the ULX as determined by \textsc{wavdetect}. At the position of the reported AGN, we use a slightly smaller radius aperture of 18\arcsec{} (so that it does not intersect the extraction region for the ULX) centered on the expected position of the AGN based on the \nustar{} positional offset.\footnote{To account for this offset, we loaded into \textsc{ds9} the \textsc{wavdetect}-determined \nustar{} aperture (listed in Table~\ref{table:photometry}) as well as the \chandra{} apertures for the AGN and ULX positions given by \citet{secrest2012}. We then manually shifted the two apertures from \citet{secrest2012} until the \chandra{} and \nustar{} ULX apertures were cospatial, and then recorded the expected position of the AGN within the \nustar{} image.} Because the source was not detected in FPMB, we loaded the FPMA regions in \textsc{ds9} on top of the FPMB image and used the centroid feature to center the ULX aperture to try and account for the positional offset between FPMA and FPMB, which we found to be $\sim4.6$\arcsec{}. Based on this shift, we then shifted the AGN aperture to its expected FPMB position. As for the background regions, in order to avoid sampling from areas of the image which overlap with \ngc4178{} or other potential sources in the FOV, we used a set of four 60\arcsec{} radius apertures spaced around \ngc4178{}.

For the \chandra{} observations of \ngc4178{} and J0851+3926, we used apertures (1.5\arcsec{} in radius) identical to those used in \citet{secrest2012} and \citet{bohn2020}, respectively. A circular aperture of 20\arcsec{} in radius was used to sample the background in a source free region of the ACIS-S detector for each observation. 

\textsc{dmextract} from \textsc{ciao} was used to extract source and background counts using the regions described above. We required a significance threshold of $3\sigma$ for a source to be considered detected. For sources with greater than 20 counts and a $3\sigma$ detection in a particular energy band, we assume the photon count distribution is Gaussian and compute the error on the counts in that band as $\sqrt{\rm{N}}$ where $\rm{N}$ is the number of counts. For non-detections ($<3\sigma$), we instead compute the $3\sigma$ (99.7\% confidence level) upper limits for the net source counts using the gross measured counts in the source and background regions following the Bayesian method of \citet{kraft1991}. Additionally, due to either weak or undetected sources, we compute the binomial no-source probability ($P_{\rm{B}}$) for each energy band \citep[see Eq. 1 in Section 3 of][]{lansbury2014}. We required a source to meet a threshold of ($P_{\rm{B}}$) $> 0.002$ in order to be considered statistically significant as opposed to being the result of a spurious background fluctuation.

\begin{table*}[ht]
{\centering
\caption{\nustar{} Photometry}
\label{table:photometry}
\begin{tabular}{ccccccc}
\hline
\hline
\noalign{\smallskip}
\noalign{\smallskip}
System & FPM & $\alpha$ & $\delta$ & \multicolumn{3}{c}{Net Counts} \\
 & & & & 3-24 keV & 3-10 keV & 10-24 keV \\
(1) & (2) & (3) & (4) & (5) & (6) & (7) \\
\noalign{\smallskip}
\noalign{\smallskip}
\hline
\noalign{\smallskip}
\ngc4178{}, ULX & A & 12\hr{}12\min{}44.\sec{}864   & +10\deg{}51\min{}20.\sec{}709 & $47.0 \pm 12.9$ &  $38.3 \pm 10.2$ & $<28.6$  \\ 
 & B & 12\hr{}12\min{}44.\sec{}588   & +10\deg{}51\min{}18.\sec{}054 & $38.8 \pm 12.5$ &  $29.7 \pm 9.8 $ & $<28.8$  \\ 
\ngc4178{}, AGN & A & 12\hr{}12\min{}46.\sec{}663   & +10\deg{}52\min{}01.\sec{}711 & $<21.2$         &  $<21.5        $ & $<10.3$  \\ 
 & B & 12\hr{}12\min{}46.\sec{}394   & +10\deg{}51\min{}58.\sec{}366 & $<29.7$         &  $<23.8        $ & $<16.0$  \\ 
J0851+3926    & A & 08\hr{}51\min{}53.\sec{}743   & +39\deg{}26\min{}11.\sec{}526 & $<62.8$         &  $<40.8$         & $<40.2$  \\ 
 & B & & & $<41.0$         &  $<33.5$         & $<29.0$  \\ 
\noalign{\smallskip}
\hline
\end{tabular}
}
\tablecomments{\nustar{} photometry for the 3-24 keV, 3-10 keV, and 10-24 keV energy bands. Columns 1-2: system name and FPM. Columns 3-4: right ascension and declination of the source apertures. Columns 5-7: net source counts for the 3-24 keV, 3-10 keV, and 10-24 keV energy bands (see Section~\ref{sec:analysis}). 
}

\end{table*}

\subsection{Flux Calculations}
For the AGNs in \ngc4178{} and J0851+3926, we used the \chandra{} \textsc{pimms} toolkit to calculate upper limits for the observed fluxes in each of the 3-24 keV, 3-10 keV, and 10-24 keV bands (as well as the 2-10 keV band for direct comparison to previous work with \chandra{}) based on the count rates derived for each position. \textsc{pimms} assumes the input is the total, two-camera (FPMA + FPMB) count rate and is retrieved from an aperture enclosing 50\% of the enclosed energy fraction (EEF), which corresponds to an aperture size of $\sim30''$ for \nustar{} \citep{harrison2013}. During this step, for each band we combined the count rate upper limits derived from FPMA and FPMB and where appropriate applied a scale factor to correct the count rates to the 50\% EEF\footnote{Thus, the count rates for NGC 4178 were scaled up to 50\% and the rates for J0851+3926 were scaled down to 50\%}. We then provided these count rates to \textsc{pimms}. We assumed a power law model with photon index $\Gamma=1.8$ \citep{mushotzky1993,ricci2017bass} for J0851+3926, while we calculated two fluxes per energy band for NGC 4178, one assuming $\Gamma=2.6$ (the unobscured AGN case from \citealp{secrest2012}) and one assuming $\Gamma=2.3$ (the obscured case from \citealp{secrest2012}). We also took into account Galactic absorption along the line of sight, obtained from the \textit{Swift} Galactic \nh{} calculator \citep{willingale2013}\footnote{https://www.swift.ac.uk/analysis/nhtot/}.

\subsection{Indirect Estimates of AGN Column Densities}

Without sufficient counts to enable direct spectral fitting, we instead estimated a lower limit on the column density along the line of sight to the AGNs in both galaxies using (1) the ratio between the observed upper limit on the 10-24 keV \nustar{} flux and the expected, intrinsic 10-24 keV flux and (2) sets of attenuation curves \citep[see, e.g., Fig. 1 in ][]{ricci2015} generated with \textsc{xspec} \citep{arnaud1996}. First, we established a model in \textsc{xspec} consisting of a primary power law, photoelectric absorption (\textsc{tbabs}) and Compton scattering (\textsc{cabs}), and a component to account for reprocessed radiation (\textsc{Borus}; \citealp{balokovic2018}), as well as a component for Thomson scattered X-ray radiation (\textsc{f*cuffpl}), expressed as: 
\begin{equation*}
    (\textsc{f}\times\textsc{cutoffpl}) +(\textsc{tbabs}\times\textsc{cabs}\times\textsc{cutoffpl})+\textsc{Borus}
\end{equation*}
which we stepped through a range of column densities $22\leq\rm{log}(\nhm{})\leq25.4$ in increments of $\Delta\rm{log}(\nhm{})=0.2$. We assumed a power law with a photon index of $\Gamma=1.8$ \citep[e.g.,][]{mushotzky1993,ricci2017bass} for J0851+3926 and we used two different photon index choices ($\Gamma=2.6$ and $\Gamma=2.3$) for \ngc4178{}, as we did in the PIMMS flux calculations. Furthermore, we assumed a scattering fraction of 0.5\% ($f=0.005$, similar to scattering fractions found for obscured \swift{} AGNs, e.g., \citealp{ricci2017bass}, and close to what might be expected in the case of heavily buried AGNs, e.g., \citealp{ueda2007}), and we tested this model using two different covering factors (C): 0.5 and 0.99. These choices reflect two scenarios: one in which the AGNs host a more `standard' obscuring torus (C=0.5) and one in which the AGN is heavily buried with a high covering factor (C=0.99). For each step through log($\nhm{}$) space, we recorded the 2-10 keV and 10-24 keV fluxes. Using these fluxes, we then built two attenuation curves, one for C=0.5 and one for C=0.99, for the hard X-ray 10-24 keV emission of both galaxies, where the curves are normalized to the `intrinsic' 10-24 keV flux measured when $\rm{log}(\nhm{})=22.0$. 

We then computed the expected, intrinsic 10-24 keV fluxes of each AGN. For J0851+3926, this involved first computing the expected 2-10 keV flux based on the \wise{} $12\rm{\,\mu m}$ fluxes \citep[using the relation from][]{asmus2015}, and then converting the intrinsic 2-10 keV flux to the intrinsic 10-24 keV flux using \textsc{XSpec}, again assuming $\Gamma=1.8$. For NGC 4178, we instead adopted the 2-10 keV fluxes found by \citep{secrest2012} using \textit{Chandra}: $6.9\times10^{38}$ erg~cm$^{-2}$~s$^{-1}$ for the unobscured case (which used a power law model with $\Gamma=2.6$), and $8.6\times10^{39}$ erg~cm$^{-2}$~s$^{-1}$ in the obscured case (which used a power law model with $\Gamma=2.3$).

Finally, we took the ratio between the observed upper limit on the 10-24 keV \nustar{} flux (Table~\ref{table:fluxes}) and the expected, intrinsic 10-24 keV flux, and used the attenuation curves to obtain lower limits on the column density along the line of sight. We thus obtain for each galaxy two lower limits on the column density, one assuming C=0.5 and one assuming C=0.99. Both lower limits are quoted in this work for J0851+3926. For NGC 4178, however, the intrinsic X-ray fluxes used in these calculations were model specific: the intrinsic flux for the obscured case was derived \textit{assuming} $\Gamma=2.3$ and a high covering factor (C=0.99), while the unobscured case assumed $\Gamma=2.6$ and had no requirements for the covering factor. To maintain consistency, for the unobscured case ($\Gamma=2.6$) we quote only the $N_{\rm{H}}$ lower limit derived using C=0.5, and for the obscured case ($\Gamma=2.3$) we quote only the the $N_{\rm{H}}$ lower limit derived using C=0.99.  

\section{Results}
\label{sec:results}
We show the \nustar{} 3-10\,keV, 10-24\,keV, and 3-24\,keV images along with Pan-STARRS r-band images of \ngc4178{} and J0851+3926 in Figure~\ref{fig:imaging}. Neither of the reported AGNs in \ngc4178{} and J0851+3926 are detected in any \nustar{} energy band, so we instead calculated the upper limits on the counts and fluxes for each AGN candidate (as outlined in Section~\ref{sec:analysis}). The ULX in \ngc4178{} was significantly detected in the 3-10 keV and 3-24 keV images. We report aperture positions and count statistics in Table~\ref{table:photometry} and X-ray fluxes of the AGNs in Table~\ref{table:fluxes}. In the case of \ngc4178{}, the non-detection suggests there is no X-ray emitting AGN with an observed flux in excess of $7.41\times10^{-14}$\,erg\,cm$^{-2}$\,s$^{-1}$ in the 10-24 keV band (assuming $\Gamma=2.3)$. For J0851+3926, the non-detection suggests there is no X-ray emitting AGN with an observed flux in excess of $9.40\times10^{-14}$\,erg\,cm$^{-2}$\,s$^{-1}$ ($\Gamma=1.8$) in the 10-24 keV band. To offer a direct comparison to previous work with \chandra{}, we also derived the 2-10 keV flux upper limits for each system using the flux in the \nustar{} 3-10 keV band: there are no X-ray emitting AGNs with observed 2-10 keV fluxes in excess of $7.36\times10^{-14}$\,erg\,cm$^{-2}$\,s$^{-1}$ in the case of \ngc4178{} ($\Gamma=2.3$) and $4.84\times10^{-14}$\,erg\,cm$^{-2}$\,s$^{-1}$ ($\Gamma=1.8$) in the case of J0851+3926. These 2-10 keV flux upper limits are consistent with the previous upper limits established using \chandra{} \citep{secrest2012,bohn2020}.

We also recalculated the 2-10\,keV upper limits on the counts and flux for J0851+3926 using the archival \chandra{} imaging and used the appropriate Bayesian approach mentioned in Section~\ref{sec:analysis}. We found the upper limit on the counts to be $<8$ counts in the 0.3-8\,keV band, $<8$ counts in the 0.3-2\,keV band, and $<6$ counts in the 2-8\,keV band based on the \chandra{} imaging. These counts in turn yield flux upper limits of $<5.30\times10^{-15}$, $<4.14\times10^{-15}$, and $<5.51\times10^{-15}$ erg\,cm$^{-2}$\,s$^{-1}$ in the 0.3-8 keV, 0.3-2 keV, and 2-8 keV \chandra{} bands, respectively. Using the 2-8\,keV flux upper limit, we found the 2-10\,keV flux upper limit to be $6.55\times10^{-15}$\,erg\,cm$^{-2}$\,s$^{-1}$. These flux upper limits supersede those calculated in \citet{bohn2020}.

In the absence of spectra, we have inferred lower limits to the column densities required for non-detections in the \nustar{} energy bands using the flux ratio calculations described in Section~\ref{sec:analysis}; we report these column densities in Table~\ref{table:fluxes}. In order to remain undetected by \nustar{}, the central source in J0851+3926 would need to be obscured by log($N_{\rm{H}}/\rm{cm}^2)>24.1$, assuming a covering factor of C=0.5, or log($N_{\rm{H}}/\rm{cm}^2)>24.3$ assuming C=0.99. If NGC\,4178 hosts a heavily obscured AGN, as posited by \citet{secrest2012}, it would need to be obscured by log($N_{\rm{H}}/\rm{cm}^2)>24.2$ (assuming $\Gamma=2.3$ and C=0.99) to remain undetected by \nustar{} in the 10-24 keV band. If \ngc4178{} instead hosts an unobscured AGN \citep[$\Gamma=2.6$,][]{secrest2012}, its intrinsic flux is below the detection limit of the \nustar{} observation.\footnote{The inferred column density, log($N_{\rm{H}}/\rm{cm}^2)=22$, is the minimum possible value using our attenuation curve method, and consistent with an unobscured AGN. In this case, log($N_{\rm{H}}/\rm{cm}^2)=22$ should be considered an \textit{upper limit} to the column density.}

\begin{table*}[t]
{\centering
\caption{\nustar{} Flux Upper Limits and Column Density Lower Limits for the AGNs}
\label{table:fluxes}
\begin{tabular}{ccccccccc}
\hline
\hline
\noalign{\smallskip}
\noalign{\smallskip}
System & $\Gamma$ & \multicolumn{4}{c}{Observed Flux} & Covering Factor & log($N_{\rm{H}}/\rm{cm}^{-2}$) \\
 & & \multicolumn{4}{c}{($10^{-14}$ erg cm$^{-2}$ s$^{-1}$)} & & \\
 & & 2-10 keV & 3-24 keV & 3-10 keV & 10-24 keV & & \\
(1) & (2) & (3) & (4) & (5) & (6) & (7) & (8) \\
\noalign{\smallskip}
\noalign{\smallskip}
\hline
\noalign{\smallskip}
\ngc4178{}    & 2.6 & $<8.03 $ & $<6.99 $ & $<5.23 $ & $<7.14 $ & 0.5 & $\leq22.0$ \\ 
\ngc4178{}    & 2.3 & $<7.36 $ & $<7.30 $ & $<5.16 $ & $<7.41 $ & 0.99 & $>24.2$ \\
J0851+3926    & 1.8 & $<4.84$ & $<7.47$ & $<3.77$ & $<9.40$     & 0.5 (0.99) & $>24.1$ ($>24.3$) \\ 
\noalign{\smallskip}
\hline
\end{tabular}
}
\tablecomments{Observed flux upper limits and column density lower limits for the candidate AGNs in \ngc4178{} and J0851+3926. Cols 1-2: system name and choice of photon index for the modeling. Cols 3-6: observed flux upper limits in units of $10^{-14}$ erg cm$^{-2}$ s$^{-1}$ in the 2-10 keV, 3-24 keV, 3-10 keV, and 10-24 keV energy bands. Col 7: covering factor choice(s) used when inferring column density limits (see Section~\ref{sec:analysis}). Col 8: inferred column densities derived from the flux ratios and attenuation curves developed in Section~\ref{sec:analysis}. Column densities are rounded to the nearest tenth.
}
\end{table*}

\section{Discussion}
\label{sec:discussion}

\subsection{\ngc4178{}}
The lack of a hard X-ray point source in \ngc4178{} offers little constraint on the presence or absence of an AGN. Based upon our analysis, if an X-ray emitting AGN exists in the nucleus of \ngc4178{}, it is either a Compton-thick AGN with log($N_{\rm{H}}/\rm{cm}^{-2})>24.2$ -- in general agreement with the column density estimated by \citet{secrest2012} for their heavily obscured AGN case -- or a low luminosity AGN (LLAGN) with column density log($N_{\rm{H}}/\rm{cm}^{-2})\leq22$ and flux level below the detection limit of \nustar{}. 

Recently, \citet{hebbar2019} examined the \chandra{} spectra of the nuclear source in \ngc4178{} and found that the spectrum is poorly fit by an absorbed power law model (\textsc{pcfabs*pegpwrlw}) -- which would arise in the case of an obscured AGN -- but is better fit and more resembles emission from a supernova remnant, modeled via a hot plasma model (\textsc{tbabs*vapec} in \textsc{xspec}). This result suggests that the X-ray emission is not dominated by an AGN either because there is no AGN or because it is too obscured. To compare the best fitting model from \citet{hebbar2019} to the non-detections in the \nustar{} imaging, we refit the \chandra{} 0.3-8 keV spectra using the \textsc{tbabs*vapec} model and froze all parameters to the best-fit values in Table~2 from \citet{hebbar2019} except for the normalization, which was left free to vary. We then fit the model to recover the approximate normalization. Using this model, we found that the absorbed hot plasma model returns\footnote{These error bounds should be considered approximate, as the bounds change during each iterative check in \textsc{Xspec} with the error command.} an observed 2-10 keV X-ray flux of $2.0^{+0.3}_{-0.4}\times10^{-15}$ erg\,cm$^{-2}$\,s$^{-1}$ and a 3-10 keV observed flux of $6.4^{+1.0}_{-1.0}\times10^{-16}$ erg\,cm$^{-2}$\,s$^{-1}$. These fluxes are a magnitude or more lower than the upper limits derived for any AGN in \ngc4178{}, placing such a source well below the detection threshold of the \nustar{} imaging. At this expected flux level a non-detection is consistent with the picture that the nuclear X-ray flux is dominated by a supernova remnant, but unfortunately the \nustar{} data are inconclusive in this respect, and cannot provide strong evidence for or against either scenario.

\begin{figure}
    \centering
    \includegraphics[width=\linewidth]{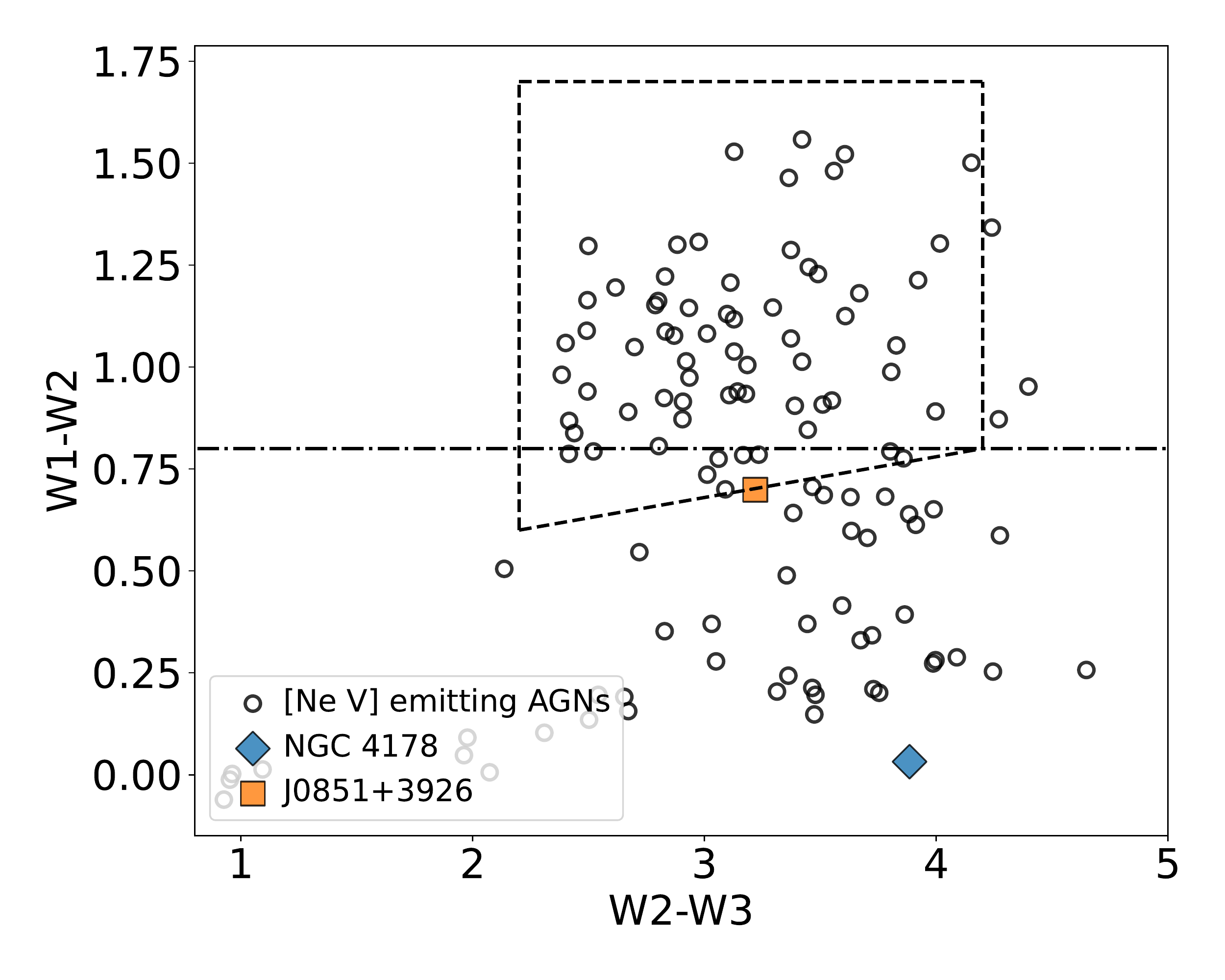}
    \caption{\wise{} color-color space for [Ne V] emitting AGNs. The \wise{} $W1-W2$ color is given on the y-axis, while the $W2-W3$ color is given on the x-axis. The horizontal dash-dotted line at $y=0.8$ denotes the mid-IR AGN criterion from \citet{stern2012}, while the dashed black wedge is the AGN criterion from \citet{jarrett2011}. [Ne V] emitting AGNs are plotted as open circles, while NGC 4178 is plotted using a blue diamond. Like many other [Ne V] emitting AGNs, NGC 4178 also does not manifest as an AGN in the mid-IR. J0851+3926 is also plotted here as an orange square.}
    \label{fig:mir_color_ngc4178}
\end{figure}

In an effort to uncover any trace of AGN activity in \ngc4178{}, we examined the SDSS optical spectra to search for optical coronal lines which might have arisen due to an AGN; no optical coronal lines are found in the optical spectra. Similarly, we have examined the \wise{} mid-IR colors from the AllWISE point source catalog \citep{https://doi.org/10.26131/irsa1}, as well as the mid-IR colors derived from the NEOWISE reactivation database \citep{https://doi.org/10.26131/irsa144}; at no point in time do we find any evidence for a mid-infrared AGN based on the commonly employed selection criteria of \citet{jarrett2011} and \citet{stern2012}. We should note that while \ngc4178{} does not manifest as an AGN in the mid-IR, its mid-IR colors are not \textit{unlike} those of other [Ne V] emitting AGNs in the local Universe; Figure~\ref{fig:mir_color_ngc4178} shows that \ngc4178{} (blue diamond) occupies the same mid-IR color space as that of other, local redshift, [Ne V] emitting AGNs (open circles) that fall outside of the selection criteria from \citet{jarrett2011} and \citet{stern2012}, so the lack of a mid-IR AGN should not immediately discount an AGN scenario. However, given the ambiguity surrounding the nature of the nuclear X-ray source and the lack of optical and mid-infrared signatures, we cannot provide any definitive evidence which supports or refutes either scenario proposed by \citet{secrest2012} and \citet{hebbar2019}.

Interestingly, this situation is reminiscent of NGC 3486 -- as one example -- in that a high excitation line attributed to AGN activity was reported \citep[O VI in NGC 3486 and Ne V in \ngc4178{},][]{annuar2020,satyapal2009} and an AGN was reported based on soft X-ray emission from \xmm{} \citep[$L_{\rm{2-10\,keV}}=1.1\times10^{39}$ erg s$^{-1}$ in NGC 3486,][]{cappi2006} or \chandra{} \citep[$L_{\rm{2-10\,keV}}=7.9\times10^{38}$ erg s$^{-1}$ in \ngc4178{},][]{secrest2012} but was not detected in the \nustar{} imaging \citep[][and this work]{annuar2020}. The major difference in this comparison is that NGC 3486 is detected as a Type II AGN in the optical, whereas \ngc4178{} presents as an HII star forming region.

Perhaps the most plausible explanation for the lack of a hard X-ray or optical AGN in \ngc4178{} is that the [\ion{Ne}{5}] detection is a light echo tracing AGN activity in the past, as has recently been suggested by \cite{graham2019}. The \textit{Spitzer} SH slit subtends a large fraction of the nuclear region of \ngc4178{} (but it does not overlap with the \chandra{} X-ray position of the ULX, ruling out the possibility that the ULX produced the [Ne V] line), so there is no way to accurately constrain the location of the [Ne V] emitting source. However, the detection of this line without a counterpart in the optical or hard X-ray bands suggests the simple explanation that the AGN is currently relatively quiet (possibly an LLAGN) but was more active at some point in the past \citep{graham2019}.

\subsection{J0851+3926}
As in the case of \ngc4178{}, while the \nustar{} non-detection in J0851+3926 does not provide confirmation of an AGN nor a precise constraint on the potential absorbing column, we can still compare the flux upper limits between observations and derive lower limits to the potential column density. The flux upper limit in the 2-10\,keV band derived from \nustar{}, $F_{\rm{2-10\,keV}}<4.84\times10^{-14}$ erg cm$^{-2}$ s$^{-1}$ (see Section~\ref{sec:analysis}), which corresponds to a 2-10\,keV X-ray luminosity of $L_{\rm{2-10\,keV}}<2.14\times10^{42}$ erg s$^{-1}$, is an order of magnitude higher than (but consistent with) both the upper limit determined with \chandra{} in \citet{bohn2020} and the updated 2-10 keV \chandra{} upper limit ($L_{\rm{2-10\,keV}}<2.90\times10^{41}$ erg s$^{-1}$) derived in Section~\ref{sec:analysis}. 

Based on the flux attenuation curves calculated in Section~\ref{sec:analysis}, we would expect a non-detection in the \nustar{} 10-24\,keV energy band for a column density of log($N_{\rm{H}}$/cm$^{2})> 24.1$. This is similar (albeit slightly smaller) to the column density lower limit calculated in \citet{bohn2020} using the ratio of the 2-10\,keV luminosity and the \wise{} $12\,\rm{\mu m}$ luminosity \citep{pfeifle2022}, which was determined to be $\sim24.4$ based on the published \chandra{} upper limit in \citet{bohn2020}. Repeating this latter calculation using the \textit{updated} 2-10\,keV flux upper limit from the \chandra{} data (Section~4) as well as the 2-10\,keV flux upper limit derived from the \nustar{} imaging, we find lower limits to the column density of log($N_{\rm{H}}$/cm$^{2})>24.3$ and log($N_{\rm{H}}$/cm$^{2})> 23.9$, respectively, when using expression (2) from \citet{pfeifle2022} for the column density derived from the ratio log($L_{2-10\,\rm{keV}}/L_{12\,\rm{\mu m}}$).\footnote{A slightly revised version of Equation (1) and (2) from \citet{pfeifle2022} was published following the publication of \citet{bohn2020}, which explains the slight difference in column density.} All three of these calculations yield column densities similar to one another, and therefore we do expect this AGN to be at least heavily obscured if not Compton-thick.

\begin{figure}[ht!]
    \centering
    \subfloat{\includegraphics[width=0.46\textwidth]{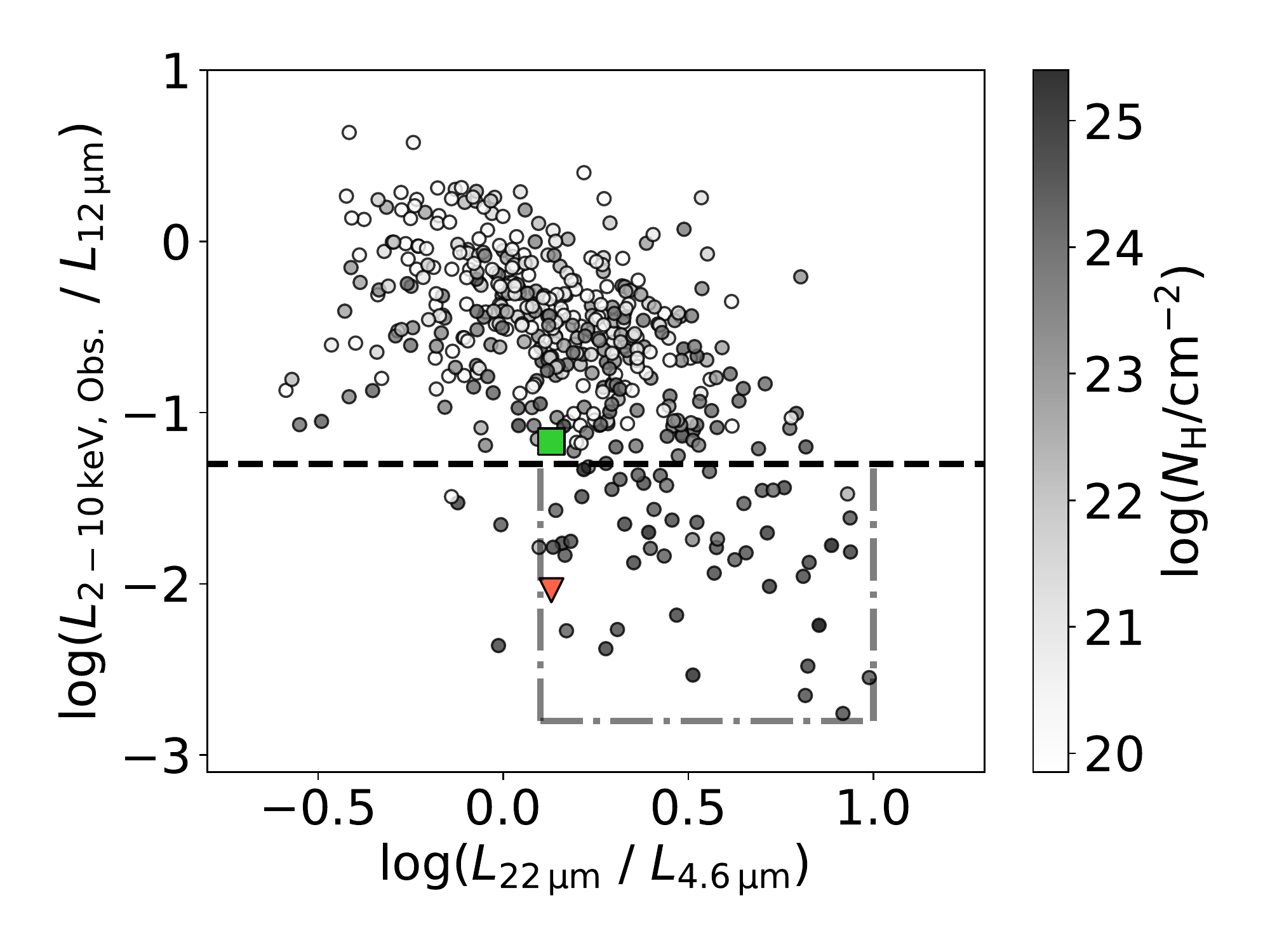}}\\
    \vspace{-4mm}
    \subfloat{\includegraphics[width=0.46\textwidth]{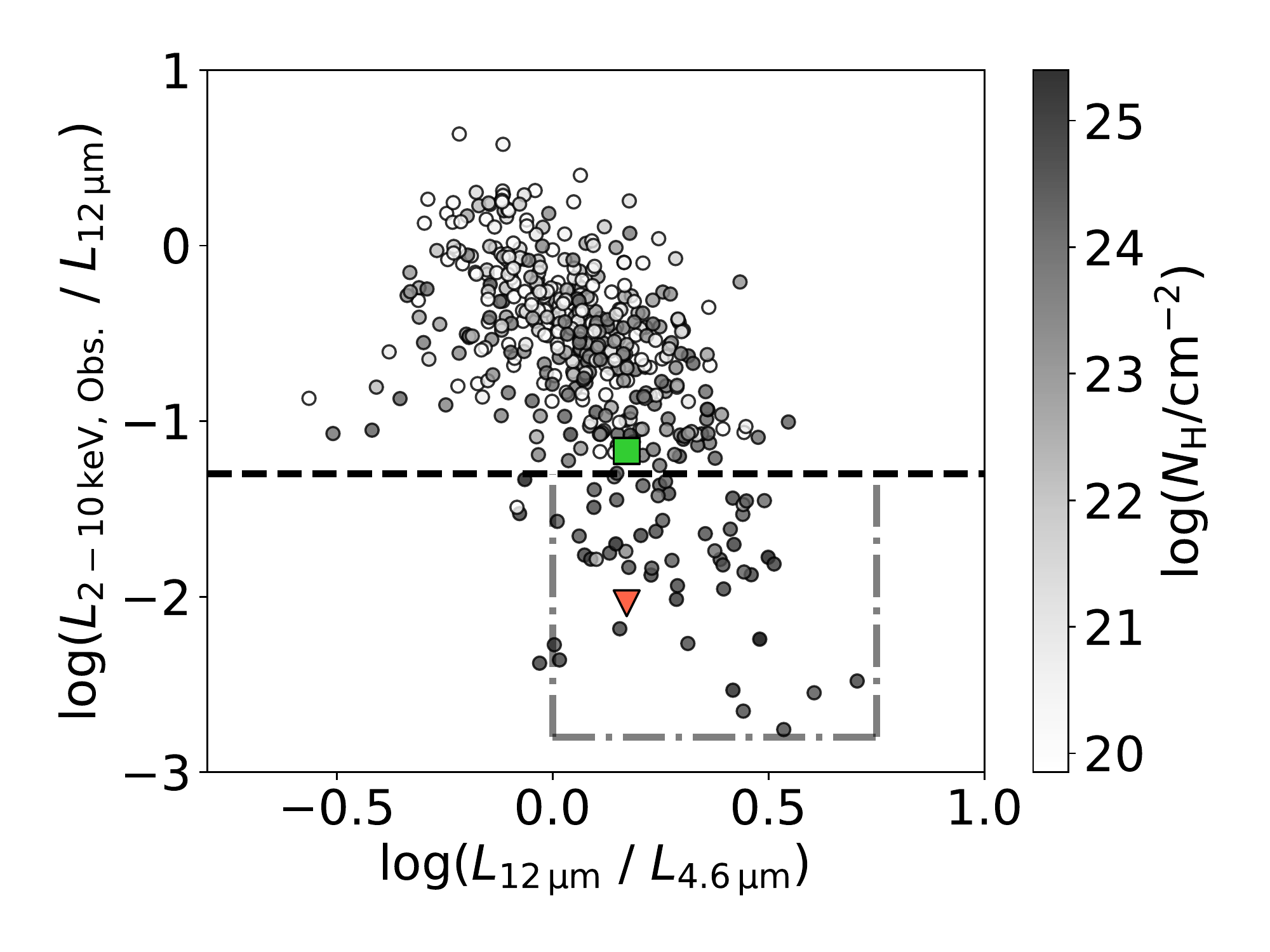}}
    \caption{Obscuration diagnostics from \citet{pfeifle2022} for J0851+3926. J0851+3926 is shown as an inverted red triangle (\chandra{}) and green square (\nustar{}) while the \swift{} AGNs are shown with a gray color map to illustrate how the AGN colors change with column density. The column densities are given on the auxiliary color map. The dashed black line in each panel shows a simple cut of log(\lx{}/\ltwel{})<-1.3 for selecting heavily obscured AGNs, while the black dashed lines in concert with the gray dashed dotted lines display diagnostic regions developed in \citet{pfeifle2022}. (Top) The log(\lfour{}/\ltwo{}) and log(\lx{}/\ltwel{}) diagnostic. (Bottom) The log(\ltwel{}/\ltwo{}) and log(\lx{}/\ltwel{}) diagnostic. In both cases, J0851+3926 would be selected as a CT AGN when using the \chandra{} flux upper limit (red), but just barely misses the cut off criteria when using the \nustar{} flux upper limit (green).
    }
    \label{fig:diagnostics}
\end{figure}

The mid-infrared colors of J0851+3926 can offer additional clues into the nature of the source and about the obscuration inherent to the system. It is important to note that J0851+3926 not only meets the stringent three-band color cut defined in \citet{jarrett2011} but also meets the color cut $W1-W2\geq0.7$ (see Figure~\ref{fig:mir_color_ngc4178}), defined by \citet{stern2012} as an 85\% reliable selection method,\footnote{But it does not meet the more commonly used 95\% reliable method of $W1-W2>0.8$ \citep{stern2012}.} as well as the 90\% reliability criterion of $W1-W2 > 0.662\times \rm{exp}\{0.232\times(W2-13.97)^2\}$ proposed by \citet{assef2013}. Given its selection by three separate and often reliable methods \citep[at least for AGNs hosted by massive galaxies, e.g.,][]{stern2012}, it would be unusual for the mid-IR emission to be driven by other dust heating sources such as star formation. It is most likely that J0851+3926 does indeed host a heavily obscured AGN that gives rise to both the mid-IR colors of the system and the broad line observed in the near-IR as reported in \citet{bohn2020}, the latter of which is compelling evidence for an AGN in its own right. 

\citet{pfeifle2022} recently published several obscuration diagnostics based on \wise{} color ratios that we can use here as another means to indirectly probe the column density. J0851+3926 exhibits mid-IR colors of log(\ltwel{}/\ltwo{})=0.17  and log(\lfour{}/\ltwo{})=0.13. Comparing these colors against the obscuration diagnostics presented in Section~3.3 of \citet{pfeifle2022}, J0851+3926 would have been flagged as a candidate CT AGN based on its log(\lfour{}/\ltwo{}) and log(\ltwel{}/\ltwo{}) colors; given that only upper limits on the 2-10 keV flux could be derived using \chandra{} and \nustar{}, it could have been selected as a candidate CT AGN based on its log(\lx{}/\ltwel{}) ratio: the more stringent 2-10 keV flux upper limit derived from \chandra{} yields log(\lx{}/\ltwel{})$<-2.04$ and the upper limit from \nustar{} yields log(\lx{}/\ltwel{})$<1.17$. We plot these diagnostic regions in Figure~\ref{fig:diagnostics} to illustrate this point; J0851+3926 is represented as a red, inverse triangle where the \chandra{} upper limit is used and as a green square where the \nustar{} upper limit is used, while the grey points indicate the \swift{} AGNs color-coded by column density. While this discussion of mid-IR colors and selection of J0851+3926 is certainly not as quantitative as the column density estimations above, it certainly is in agreement with the scenario in which J0851+3926 hosts a heavily obscured AGN with column density on the order of, or in excess of, $10^{24}\,\rm{cm}^{-2}$.

\section{Conclusion}
\label{sec:conclusion}
We have presented new \nustar{} observations of two bulgeless galaxies, \ngc4178{} and J0851+3926, in an effort to confirm the existence of the heavily buried AGNs reported by \citet{secrest2012} and \citet{bohn2020}, respectively, and better constrain their column densities. We summarize our results here: 
\begin{itemize}
    \item Neither of the AGNs in \ngc4178{} or J0851+3926 are significantly detected by \nustar{} in any of the 3-24\,keV, 3-10\,keV, or 10-24\,keV energy bands. 
    \item There are no hard X-ray emitting AGNs above an observed 10-24 keV flux limit of $7.41\times10^{-14}$ erg\,cm$^{-2}$\,s$^{-1}$ ($\Gamma=2.3$) in \ngc4178{} and $9.40\times10^{-14}$ erg\,cm$^{-2}$\,s$^{-1}$ ($\Gamma=1.8$) in J0851+3926. 
    \item The non-detections with \textit{NuSTAR} imply column densities of log($N_{\rm{H}}$/cm$^{2})>24.2$ for \ngc4178{} \citep[assuming the obscured model with $\Gamma=2.3$ and C=0.99 from][]{secrest2012} and log($N_{\rm{H}}$/cm$^{2})>24.3$ for J0851+3926 (assuming $\Gamma=1.8$ and C=0.99). If the AGN in \ngc4178{} is unobscured \citep[$\Gamma=2.6$;][]{secrest2012}, its flux is below the detection limit of the \nustar{} observation.
    \item Comparing our \nustar{} observations to the results of \citet{hebbar2019}, if a supernova is indeed responsible for the observed nuclear X-ray emission in \ngc4178{} rather than an AGN, its expected observed X-ray luminosity ($L_{\rm{2-10\,keV}}=2.0^{+0.3}_{-0.4}\times10^{-15}$ or $L_{\rm{3-10\,keV}}=6.4^{+1.0}_{-1.0}\times10^{-16}$ erg\,cm$^{-2}$\,s$^{-1}$) is well below the detection limit of our observations.
    \item J0851+3926 is most plausibly a heavily obscured AGN. \ngc4178{} could be a heavily obscured AGN, but is also plausibly a LLAGN with a flux below the detection limit of \nustar{}; the previously detected [Ne V] emission line is likely a light echo, tracing past activity of the intermediate mass black hole \citep{graham2019}.
\end{itemize}

Unfortunately, the \nustar{} observations cannot offer definitive proof of the AGNs in J0851+3926 or \ngc4178{}, but the upper limits on the fluxes and lower limits on the column densities suggest that future works should focus on near-IR imaging or spectroscopy in order to more easily peer through the presumed high column densities. This is most significantly demonstrated by \citet{bohn2020}, who found clear signatures of the AGN in J0851+3926 based on the detection of a hidden broad line region while the AGN remains elusive at both optical and (soft and hard) X-ray wavelengths. Infrared observations of these galaxies and other optically elusive AGNs with JWST may offer the most promising avenue for detecting these AGNs via high excitation coronal line emission and optically-hidden BLRs.

\acknowledgments
We would like to thank the anonymous referee for their timely and helpful review which improved the manuscript. We thank George Lansbury for sharing his Python script for calculating Bayesian upper limits. We also thank Kimberly Weaver and Koji Mukai for helpful discussions on using PIMMS for \nustar{} flux estimations. 

R. W. P. gratefully acknowledges support from NASA grant 80NSSC21K0063 and support through an appointment to the NASA Postdoctoral Program at Goddard Space Flight Center, administered by ORAU through a contract with NASA. C. R. acknowledges support from the Fondecyt Iniciacion grant 11190831 and ANID BASAL project FB210003. T. B. and G. C. acknowledge partial support by the National Science Foundation, under grant No. AST 1817233. M. R. gratefully acknowledges support from the National Science Foundation Graduate Research Fellowship under Grant No. 2141064.

This work makes use of data from the \textit{NuSTAR} mission, a project led by Caltech, managed by the Jet Propulsion Laboratory, and funded by NASA. We thank the \textit{NuSTAR} Operations, Software, and Calibration teams for their support with the execution and analysis of these observations. This research has made use of the \textit{NuSTAR Data} Analysis Software, jointly developed by the ASI Science Data Center (Italy) and Caltech. The scientific results reported in this article are based in part on data obtained from the \textit{Chandra} Data Archive and published previously in cited articles. This research has made use of software provided by the \textit{Chandra} X-ray Center (CXC) in the application packages \textsc{CIAO}. This work made use of imaging from the Pan-STARRS1 Surveys. This research has made use of the NASA/IPAC Infrared Science Archive, which is funded by the National Aeronautics and Space Administration and operated by the California Institute of Technology. This publication makes use of data products from the Wide-field Infrared Survey Explorer, which is a joint project of the University of California, Los Angeles, and the Jet Propulsion Laboratory/California Institute of Technology, funded by the National Aeronautics and Space Administration. This publication also makes use of data products from NEOWISE, which is a project of the Jet Propulsion Laboratory/California Institute of Technology, funded by the Planetary Science Division of the National Aeronautics and Space Administration.

\vspace{5mm}
\facilities{\chandra{}, \nustar{}, \wise{}, Pan-STARRS}
\software{APLpy \citep{robitaille2012}, pandas \citep{mckinney2010}, NumPy \citep{oliphant2006,walt2011,harris2020array}, SciPy \citep{virtanen2020}, \textsc{heasoft} \citep{heasoft}, \textsc{matplotlib} \citep{hunter2007}, \textsc{xspec} \citep{arnaud1996}, \textsc{pimms}, \textsc{ciao} \citep{fruscione2006}, \textsc{nustardas}, \textsc{ds9} \citep{joyce2003}}\\


\bibliography{references}{}
\bibliographystyle{aasjournal}

\end{document}